\documentclass[aps,prd,preprint,floatfix,nofootinbib,showpacs]{revtex4-1}
\usepackage{hyperref,amssymb}
\usepackage{graphicx,color,epstopdf,amsmath}

\newcommand{\fr}[2]{{\hbox{$ #1 \over #2 $}}}

\begin{document}
\preprint{IPMU 18-0068}
\bigskip

\title{$CP$ violating mode of the stoponium decay into $Zh$}
\author{ Kingman Cheung$^{1,2,3}$, Wai-Yee Keung$^{4}$, Po-Yan Tseng$^{5}$}
\affiliation{
$^1$ National Center for Theoretical Sciences, Hsinchu,
Taiwan \\
$^2$Department of Physics, National Tsing Hua University, 
Hsinchu 300, Taiwan \\
$^3$Division of Quantum Phases \& Devices, School of Physics, 
Konkuk University, Seoul 143-701, Republic of Korea \\
$^4$Department of Physics, University of Illinois at Chicago, 
Illinois 60607 USA \\
$^5$Kavli IPMU (WPI), UTIAS, The University of Tokyo, 
Kashiwa, Chiba 277-8583, Japan
}

\renewcommand{\thefootnote}{\arabic{footnote}}
\date{April 12, 2018}

\begin{abstract}
We show that a novel decay mode $Zh$ of the bound state of stop-anti-stop pair 
in the ground state 
$^1S_0(\widetilde t_1\widetilde t_1^*)$ may have a significant
branching ratio if the $CP$ violating mixing appears in the stop
sector, even after we apply the stringent constraint from the measurement
of the electric dipole moment (EDM) of the electron. 
We show that the branching ratio can be as large as 10\%
in some parameter space that it may be detectable at the LHC.
\end{abstract}

\maketitle

\section{Introduction}
So far the Higgs boson discovered in 2012 is the only fundamental
particle of scalar in nature \cite{higgs}. On the other hand,
colored scalar bosons are definitely signs of physics beyond the 
standard model (SM), which often appears in many new physics models.
One outstanding example is the scalar-top (stop) quark -- superpartner of
the top quark --  in the
Minimal Supersymmetric Standard Model (MSSM). Their strong interaction
allows them to be produced abundantly at hadron colliders if 
kinematically allowed. The current search for the stop at LHC has
pushed its mass above about 500 GeV \cite{lhc-stop}.
%o
To escape its detection, the
mass of the lightest stop state $\widetilde t_1$ is compressed just above
the lightest neutralino mass so that there is not much missing momentum
for tagging the event at the LHC.  In such a scenario, 
the stop state is rather long lived in
comparison to the time scale of QCD hadronization.
Therefore, the stop-anti-stop pair can form the bound state, called
the stoponium\cite{Barger:1988sp}, 
which is produced through the gluon-gluon fusion
\cite{Barger:1988sp,Inazawa:1993xu,Drees:1993yr}
as in the squarkonium\cite{Herrero:1987df} production.
The ground state 
$\widetilde\eta \equiv \ ^1S_0(\widetilde t_1 \widetilde t_1^*)$ 
of the stoponium can then be identified by its
distinctive decay modes, such as $hh$, $WW$, $ZZ$, $\gamma\gamma$,
etc. Among them, the channel $hh$ stands out\cite{Barger:1988sp} 
for its significant decay
rate with clean detection signature. 
Recent studies of the stoponium at LHC  can be found in
\cite{Martin:2008sv,Barger:2011jt,Kim:2014yaa,Kumar:2014bca,Batell:2015zla,Duan:2017zar}.
There are also efforts in studying the QCD corrections
\cite{Martin:2009dj,Younkin:2009zn}, 
the lattice calculation\cite{Kim:2015zqa}, the mixing between 
the Higgs boson and the stoponium\cite{Bodwin:2016whr}
, and the role of the stoponium\cite{Keung:2017kot} in the dark matter 
co-annihilation.

Surprisingly in all studies about the stoponium decay, the channel
$Zh$ is not given. In fact, the process is forbidden by the underlying
assumption of  the $CP$ conservation, which implies  the cancellation of 
amplitudes  \'a la the Furry theorem.
However, there is no strong
argument against $CP$ violation in the stop sector.  We are going to
show in this article that $\widetilde\eta \to hZ$ can have a 
significant
branching ratio when $CP$ violating parameters are chosen yet within the
experimental constraint due to the electron electric dipole moment
(eEDM) measurement.

If the mass of the stoponium is close to the mass $m_A$ 
of the pseudoscalar Higgs boson, substantial enhancement of the 
$Zh$ decay mode happens due to the resonance effect.
Neverthless, for a stoponium mass around 1.2 TeV $\sim m_A$ the eEDM
places a very stringent constraint on the choice of the $CP$-violating
parameter such that the $B(\widetilde{\eta} \to Zh) \sim 10^{-3}$.
On the other hand, if the mass of the second stop is not too far 
from the lightest stop, substantial cancellation between the stop
contributions to the eEDM can happen, such that the $CP$-violating
parameter can be chosen to be much larger and the 
branching ratio $B(\widetilde{\eta} \to Zh) \sim 10^{-1}$.
In the extreme case that the $m_A \to \infty$ when the eEDM is not effective, 
the branching ratio can reach a large value, 
$B(\widetilde{\eta} \to Zh) \sim O(0.5)$.
This is the major result of our work.
Furthermore, due to the heavy stoponium decay the $Z$ and $h$ bosons
are very boosted, in which both bosons can be identified as boosted 
objects with advanced boost techniques to suppress backgrounds.
Such rather straightforward detection of the $Z$ and $h$ bosons makes 
the mode $Zh$ a wonderful place to look for the new 
particle as well as $CP$ violation.

The organization is as follows. In the next section, we give details about
the mixing in the stop sector, as well as the $CP$-violating couplings
to the Higgs boson and $Z$ boson. In Sec. III, we analyze the decay
mode $Zh$ together with the eEDM constraint. 
In Sec. IV, we estimate the observability of the $Zh$ mode at the LHC.
We summarize in Sec. V.

\section{$CP$-violation in the stop sector}
Let us start with the $Z$ boson couplings to the  stops
$t_i(i=1,2)$. The convective current among stop states is 
\def\partials{{\ }_\partial^\leftrightarrow}
\def\partials{\stackrel{\leftrightarrow}{\partial}}

$$ J_{ij}^\mu=i \widetilde t_i^* \partials \widetilde t_j 
\quad {\rm where }\quad 
 \partials  \equiv  \stackrel{\rightarrow}{\partial} 
                 -  \stackrel{\leftarrow}{\partial}  \ .$$
Our convention for the Feynman vertex amplitude is
$$  \langle \widetilde t_i (p_i) | J_{ij}^\mu | \widetilde t_j  (p_j) \rangle
=(p_j + p_i)^\mu \ , $$
for the incoming $p_j$ and the outgoing $p_i$.
Under the charge conjugation $C$, 
$ \widetilde t_i \stackrel{C}{\longleftrightarrow}  \widetilde t_i^* $. So
$ J_{ij}^\mu   \stackrel{C}{\longleftrightarrow}  -J_{ji}^\mu  $. 
The negative sign in the transformation of $J$ 
comes from that in $\partials$.  Consequently, we
need to make the $C$-odd transformation for the $Z$ gauge boson,
$Z^\mu \stackrel{C}{\longleftrightarrow} -Z^\mu$.
The hermiticity of the unitary interaction 
$ {\cal L} \supset  \sum_{ij} g^Z_{ij} J_{ij}^\mu Z_\mu    $
requires {$g^Z_{ij}=g^{Z*}_{ji}$}. 
If the charge conjugation is a good symmetry, we have
{$g^Z_{ij}=g^Z_{ji}$}. 
From this, we know that a complex $g^Z_{ij}$ (for $i\ne j$)
if its phase is not removable implies $C$-parity violation

In general, if the states $\widetilde t_{L,R}$ mix with each other 
by the complex $2\times2$ matrix into the mass eigenstates 
$\widetilde t_{1,2}$, 
we expect the complex off-diagonal  
$g^Z_{12}$ coupling to the $Z$ boson. 
However, we can set $g^Z_{12}$ real by
redefining the relative phase between the two stop fields $\widetilde
t_1, \widetilde t_2$. Indeed in the next
section, we adopt such a choice in our convention. To have a genuine
$C$-parity non-conservation, we need additional complex coupling coefficient 
$y$,
which appears in the Higgs vertex of $y h(\widetilde t_2^*  \widetilde t_1)$.
Then there is no more freedom to remove its phase.

\begin{figure}[h!]
\qquad
\includegraphics[height=1.3in,angle=0]{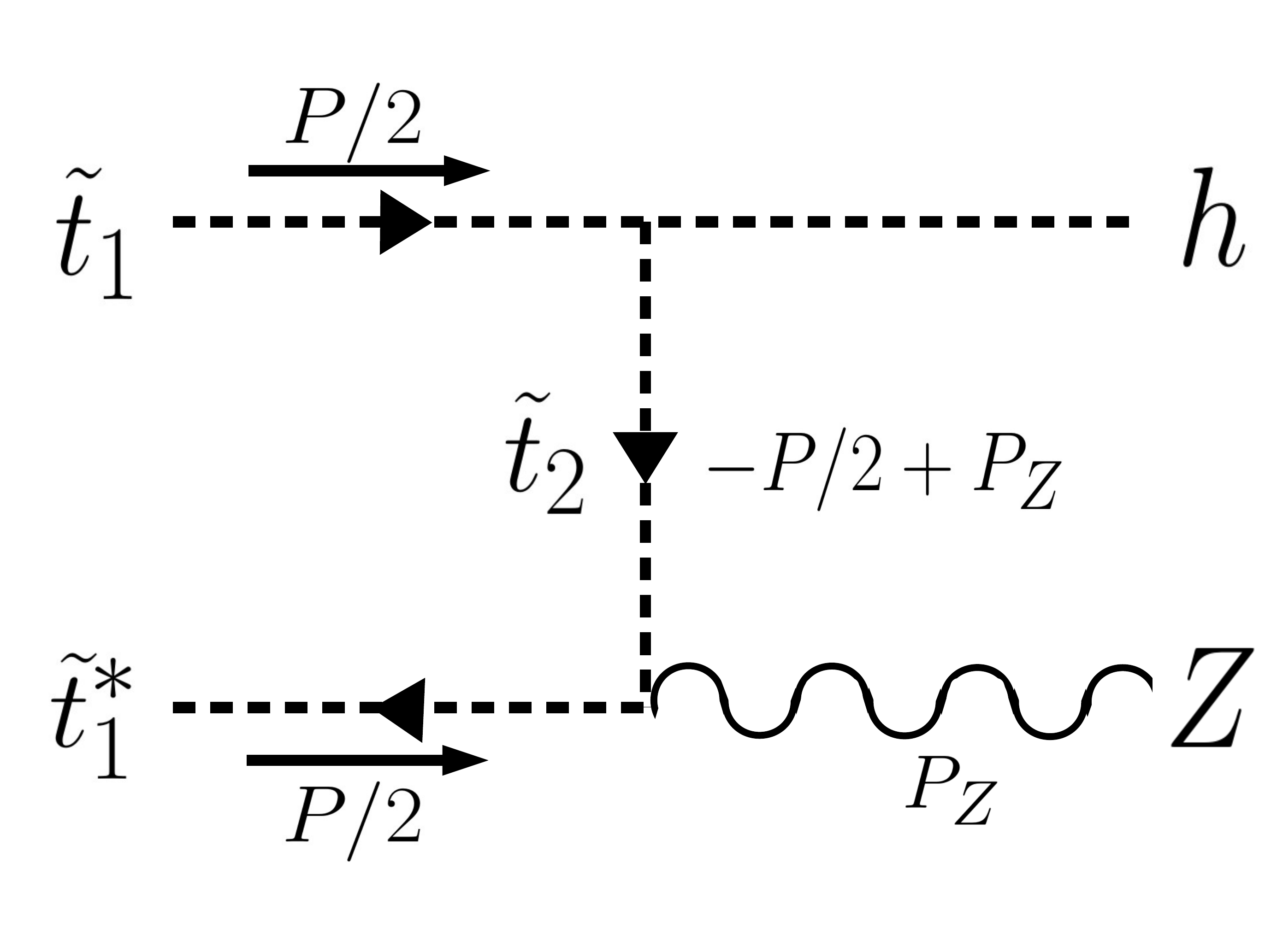}
\ 
\includegraphics[height=1.3in,angle=0]{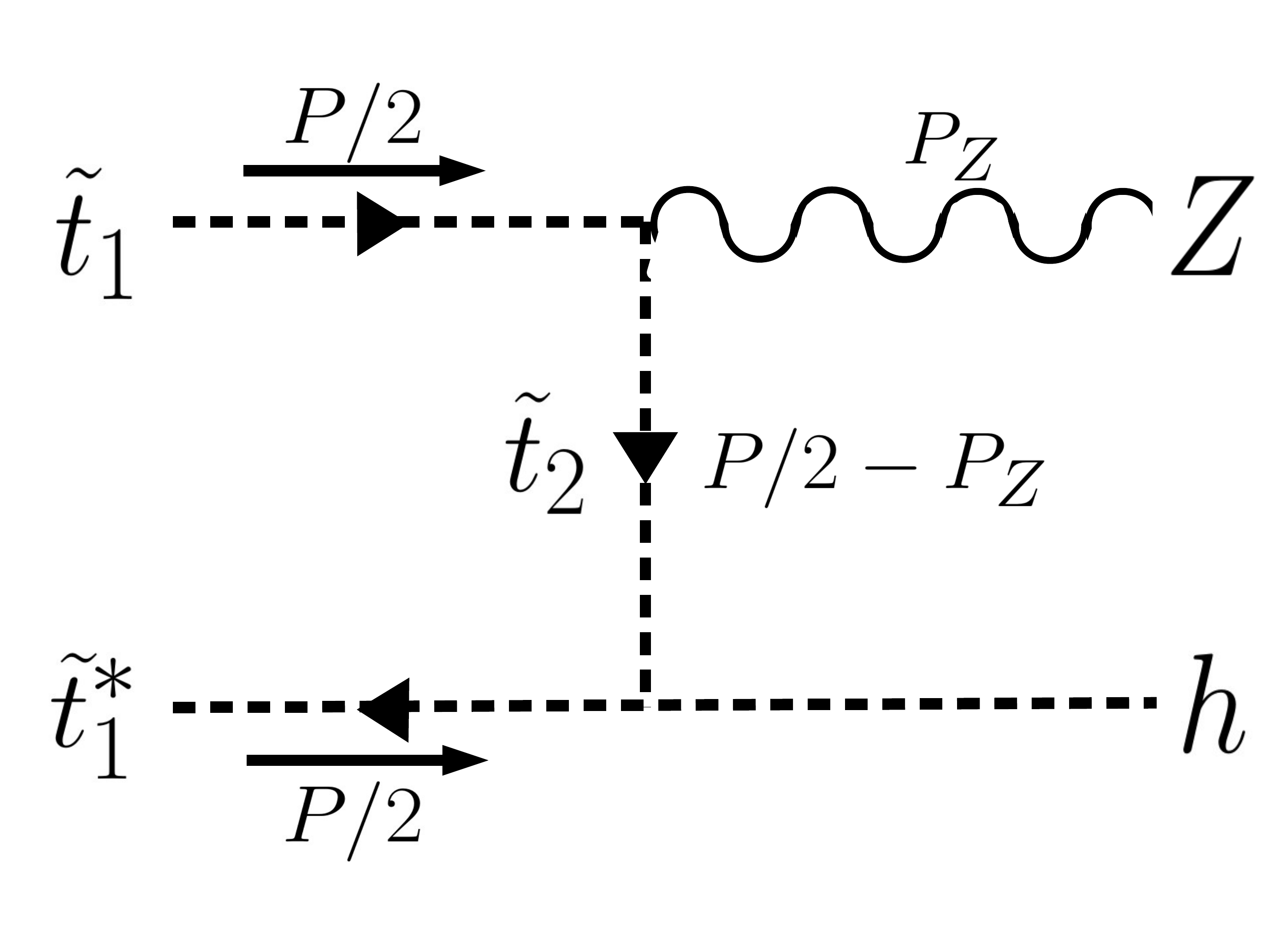}
\ 
\includegraphics[height=1.3in,angle=0]{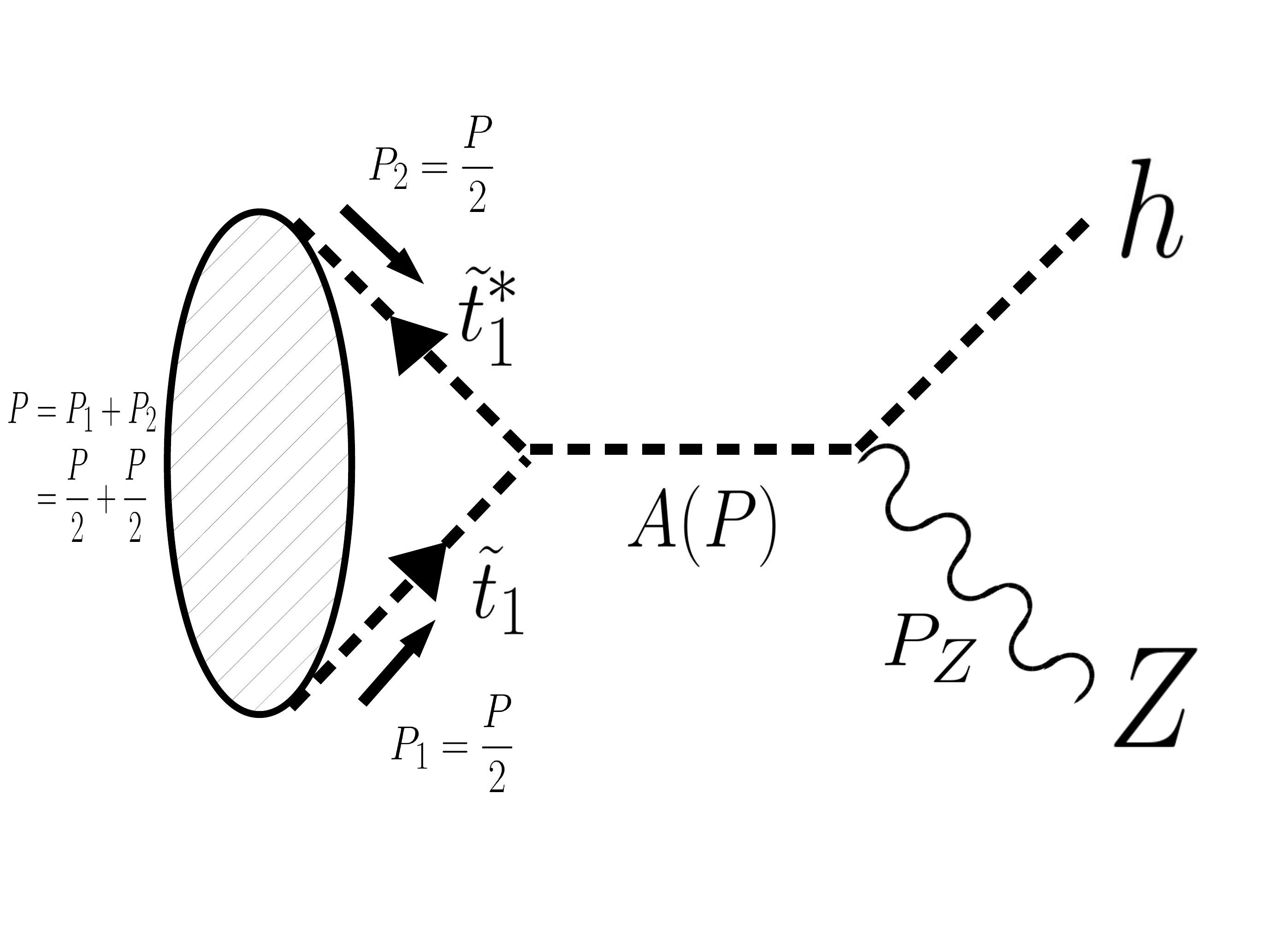}
\caption{\label{utsDiagram}
Feynman diagrams for the stoponium decaying into $Zh$ via the 
$t$,$u$,$s$ channels from the left to the right.}
\end{figure}

For the renormalizable interaction of the pure bosonic sector, 
operators of dim 4 or less do not involve the $P$-odd 
Levi-Civita $\epsilon$-symbol.
Therefore, the  $P$-parity is conserved in the $Z$ vertex. 
Consequently, the $C$-parity violation  is the $CP$-violation.
Our example is the decay of the ground state of the stoponium in
$^1S_0 (\widetilde t_1 \widetilde t_1^*)$ into $Zh$.  
The exchange of
$\widetilde t_2$ can appear in the $t$-channel and in the
$u$-channel, as shown in the first two diagrams in Fig.~\ref{utsDiagram}.
The phase of $g^Z_{ij}$ is tied with another vertex
$y h\widetilde t_1^*  \widetilde t_2$, and thus overall unremovable.
The two amplitudes of the $u$ and $t$ channels cancel 
if the coupling factor is real, but add up if imaginary. 
The production of $Zh$ from such a
decay is a sign of $CP$-violation. 

Furthermore, there exists the direct coupling of the pseudoscalar
$A^0$ to the stops,
$ A^0 (\widetilde t_1^*  \widetilde t_1-\widetilde t_2^*  \widetilde t_2)$,
which is $CP$-violating.  The ground state of the stoponium
$\widetilde\eta$ can annihilate into the virtual $A^0$ in the
$s$-channel as in the third diagram in Fig.~\ref{utsDiagram},
and then become $Zh$ via the $ZA^0h$ gauge vertex.
If the mass of the stoponium is close to the mass $m_A$ 
of the pseudoscalar Higgs boson, substantial enhancement of the 
$Zh$ decay mode happens, indeed the $Zh$ mode is significant 
in such a scenario.
Nevertheless, it is restricted by the eEDM especially when the 
mass eignestates of the stop sector is widely separated and $m_A$
is moderate.
When $m_A$ is chosen to be very heavy, then the constraint of eEDM
disappears and the $CP$ parameter can be chosen very large and the
branching ratio into $Zh$ can be as large as $O(0.5)$.

\subsection{Complex mixing in the Stop sector}

Input parameters in the calculation of the
$\widetilde{\eta} \to Zh$ decay mode include
masses $m_{\widetilde{t_1}}$, $m_{\widetilde{t_2}}$,
mixing parameters $\theta_{\widetilde{t}}$, $\delta_u$, 
${\rm Re}[\mu^*e^{-i\delta_u}]$, ${\rm Im}[\mu^*e^{-i\delta_u}]$
, and $\tan\beta$ is the ratio of the VEV of the two Higgs doublet.

The relative phase between the $\mu$ parameter and the trilinear $A_t$
parameter can be established in the following $\widetilde{t_L} \widetilde{t_R}^*$ 
term in the Lagrangian:
\begin{equation}
\mathcal{L} \supset  -y_t A_t \widetilde{t_L}\widetilde{t^*_R}H^0_u
+y_t\mu^* \widetilde{t_L}\widetilde{t^*_R}{H^0_d}^* + H.c.+.... \,,
\end{equation}
where $y_t=\frac{\sqrt{2}m_t}{v \sin\beta}$, $v=246$ GeV, and
\begin{eqnarray}
H^0_u &=& \frac{1}{\sqrt{2}}
\left[ v s_\beta
      +(c_\alpha h +s_\alpha H)
      +i(A^0 c_\beta-G^0 s_\beta) \right]\,, \nonumber \\
H^0_d &=& \frac{1}{\sqrt{2}}
\left[ v c_\beta
      +(-s_\alpha h +c_\alpha H)
      +i(A^0 s_\beta+G^0 c_\beta) \right]\,,
\end{eqnarray}
where $c_\beta, s_\beta$ are shorthand notation for $\cos\beta$ and $\sin\beta$,
$c_\alpha, s_\alpha$ are for $\cos\alpha$ and $\sin\alpha$, respectively,
$\tan\beta \equiv v_u / v_d$ is the ratio of the VEV of the two Higgs doublet,
and $\alpha$ is the mixing angle between the two neutral components of the
Higgs doublets.

The stop mass matrix can be expressed as
\begin{eqnarray}
(\widetilde{t^*_L},\widetilde{t^*_R})\left(\begin{array}{cc}
    m^2_t+M^2_{\widetilde{Q}}+m^2_Z(\frac{1}{2}-\frac{2}{3}x_W)\cos(2\beta)~~~~ & 
     m_t(A^*_t-\mu \cot\beta) \\[1mm]
     m_t(A_t-\mu^* \cot\beta)~~~~ & 
      m^2_t+M^2_{\widetilde{U}}+m^2_Z(\frac{2}{3}x_W)\cos(2\beta)       
             \end{array}\right)
     \left(\begin{array}{c}
     \widetilde{t_L}        \\[1mm]
     \widetilde{t_R}       
             \end{array}\right)\,.   \nonumber
\end{eqnarray}
We can define a phase $\delta_u$ by
\begin{equation}
A_t-\mu^* \cot\beta=|A_t-\mu^* \cot\beta|e^{i\delta_u}\,,
\end{equation}
then the mass matrix can be diagonalized by an orthogonal transformation
with an angle $\theta_{\widetilde{t}}$ into mass eigenstates $\widetilde{t}_1$ and 
$\widetilde{t}_2$:
\begin{eqnarray}
\left(\begin{array}{c}
 \widetilde{t_L} \\[1mm]
    \widetilde{t_R}    
             \end{array}\right)
=
\left(\begin{array}{cc}
  1~~~~ & 
     0 \\[1mm]
     0~~~~ & 
      e^{i\delta_u}     
             \end{array}\right)
\left(\begin{array}{cc}
  \cos{\theta_{\widetilde{t}}}~~~~ & 
     -\sin{\theta_{\widetilde{t}}} \\[1mm]
     \sin{\theta_{\widetilde{t}}}~~~~ & 
      \cos{\theta_{\widetilde{t}}}      
             \end{array}\right)
     \left(\begin{array}{c}
     \widetilde{t_1}        \\[1mm]
     \widetilde{t_2}       
             \end{array}\right)\,.   \nonumber
\end{eqnarray}
The stop mass matrix can be re-expressed in terms of 
$m_{\widetilde{t}_1}$, $m_{\widetilde{t}_2}$, $\theta_{\widetilde{t}}$, 
and $\delta_u$ as 
\begin{eqnarray}
(\widetilde{t^*_L},\widetilde{t^*_R})\left(\begin{array}{cc}
     m^2_{\widetilde{t}_1}\cos^2\theta_{\widetilde{t}}+m^2_{\widetilde{t}_2}\sin^2\theta_{\widetilde{t}}~~~~ & 
     e^{-i\delta_u}(m^2_{\widetilde{t}_1}-m^2_{\widetilde{t}_2})\sin\theta_{\widetilde{t}}\cos\theta_{\widetilde{t}}       \\[1mm]
     e^{i\delta_u}(m^2_{\widetilde{t}_1}-m^2_{\widetilde{t}_2})\sin\theta_{\widetilde{t}}\cos\theta_{\widetilde{t}}~~~~ & 
      m^2_{\widetilde{t}_1}\sin^2\theta_{\widetilde{t}}+m^2_{\widetilde{t}_2}\cos^2\theta_{\widetilde{t}}       
             \end{array}\right)
     \left(\begin{array}{c}
     \widetilde{t_L}        \\[1mm]
     \widetilde{t_R}       
             \end{array}\right)\,.   \nonumber
\end{eqnarray}
By comparing the off-diagonal elements of the above two stop mass matrix, 
we can express $A_t$ in terms of 
${\rm Re}[\mu^*e^{-i\delta_u}]$, and ${\rm Im}[\mu^*e^{-i\delta_u}]$:
\begin{eqnarray}
{\rm Re}[A_t e^{-i\delta_u}]&=& 
{\rm Re}[\mu^*e^{-i\delta_u}] \cot\beta
+\left(\frac{m^2_{\widetilde{t_1}}-m^2_{\widetilde{t_2}}}{m_t}\right)
\sin\theta_{\widetilde{t}}\cos \theta_{\widetilde{t}}\,, 
\nonumber \\
{\rm Im}[A_t e^{-i\delta_u}]&=&{\rm Im}[\mu^*e^{-i\delta_u}] \cot\beta \,.
\end{eqnarray}

\subsection{Relevant Couplings for $Zh$ decay mode}

The interaction between $h$ and $\widetilde{t}_{L,R}$ is
\begin{eqnarray}
\mathcal{L}&\subset&
h(\widetilde{t^*_L},\widetilde{t^*_R})\left(\begin{array}{cc}
     V_{LL}~~~~ & 
     V^*_{LR}     \\[1mm]
     V_{LR}~~~~ & 
     V_{RR}
             \end{array}\right)
     \left(\begin{array}{c}
     \widetilde{t_L}        \\[1mm]
     \widetilde{t_R}       
             \end{array}\right)\,   \nonumber \\
&=&
h(\widetilde{t^*_L},\widetilde{t^*_R})\left(\begin{array}{cc}
     -\frac{gm^2_t c_\alpha}{m_W s_\beta}
     +\frac{gm_Z}{\sqrt{1-x_W}}(\frac{1}{2}-\frac{2}{3}x_W) s_{\alpha+\beta}~~~~ & 
     -\frac{1}{2}\frac{gm_t}{m_W s_\beta}(A^*_t c_\alpha+\mu s_\alpha)     \\[1mm]
     -\frac{1}{2}\frac{gm_t}{m_W s_\beta}(A_t c_\alpha+\mu^* s_\alpha)~~~~ & 
     -\frac{gm^2_t c_\alpha}{m_W s_\beta} 
     +\frac{gm_Z}{\sqrt{1-x_W}}(\frac{2}{3}x_W) s_{\alpha+\beta}
             \end{array}\right)
     \left(\begin{array}{c}
     \widetilde{t_L}        \\[1mm]
     \widetilde{t_R}       
             \end{array}\right)\,   \nonumber \\
&\equiv&h(\widetilde{t^*_1},\widetilde{t^*_2})\left(\begin{array}{cc}
     y^h_{\widetilde{t_1}\widetilde{t_1}}~~~~ & 
     {y^h_{\widetilde{t_1}\widetilde{t_2}}}^*     \\[1mm]
     y^h_{\widetilde{t_1}\widetilde{t_2}}~~~~ & 
     y^h_{\widetilde{t_2}\widetilde{t_2}}
             \end{array}\right)
     \left(\begin{array}{c}
     \widetilde{t_1}        \\[1mm]
     \widetilde{t_2}       
             \end{array}\right)\,,
\end{eqnarray}
where $m_W=\frac{g}{2}v$, $m_Z=\frac{1}{2}\sqrt{g^2+g'^2}v$, 
$m_Z={m_W}/{\sqrt{1-x_W}}$, and
\begin{eqnarray}
y^h_{\widetilde{t_1}\widetilde{t_1}} &=& 
V_{LL}c^2_{\theta_{\widetilde{t}}}+V_{RR}s^2_{\theta_{\widetilde{t}}}
+2s_{\theta_{\widetilde{t}}}c_{\theta_{\widetilde{t}}} 
{\rm Re}[V_{LR}e^{-i\delta_u}]     \nonumber \\
y^h_{\widetilde{t_2}\widetilde{t_2}} &=&  
V_{LL}s^2_{\theta_{\widetilde{t}}}+V_{RR}c^2_{\theta_{\widetilde{t}}}
-2s_{\theta_{\widetilde{t}}}c_{\theta_{\widetilde{t}}} 
{\rm Re}[V_{LR}e^{-i\delta_u}]     \nonumber \\
y^h_{\widetilde{t_1}\widetilde{t_2}} &=& 
s_{\theta_{\widetilde{t}}} c_{\theta_{\widetilde{t}}}(V_{RR}-V_{LL})
+(c^2_{\theta_{\widetilde{t}}}-s^2_{\theta_{\widetilde{t}}}){\rm Re}[V_{LR}e^{-i\delta_u}]
+i{\rm Im}[V_{LR}e^{-i\delta_u}]  \,,
\end{eqnarray}
and
\begin{eqnarray}
{\rm Re}[V_{LR}e^{-i\delta_u}]
&=&-\frac{1}{2}\frac{gm_t}{m_W}
\left\lbrace
\frac{\cos(\beta-\alpha)}{s^2_\beta}
{\rm Re}[\mu^* e^{-i\delta_u}]
+\frac{c_\alpha}{s_\beta}\left(\frac{m^2_{\widetilde{t_1}}-m^2_{\widetilde{t_2}}}{m_t}\right) \sin{\theta_{\widetilde{t}}}\cos{\theta_{\widetilde{t}}}
\right\rbrace\,, \nonumber \\
{\rm Im}[V_{LR}e^{-i\delta_u}]
&=&-\frac{1}{2}\frac{gm_t}{m_W}
\left\lbrace
\frac{\cos(\beta-\alpha)}{s^2_\beta}
{\rm Im}[\mu^* e^{-i\delta_u}]
\right\rbrace\,.   \label{7}
\end{eqnarray}
For the interaction between the heavy Higgs $H$ and stops 
$\widetilde{t}_{1,2}$,
we need to change the above  $h$, $\widetilde{t}_{1,2}$ interactions by
substitutions
\begin{equation}
h       \longrightarrow H \ ,\
c_\alpha \longrightarrow s_\alpha  \ ,\
-s_\alpha  \longrightarrow c_{\alpha}  \ .
\end{equation}

On the other hand, 
the interaction between $A^0$ and $\widetilde{t}_{L,R}$ is
\begin{eqnarray}
\mathcal{L}& \supset&
-\frac{im_t}{v\sin\beta}A^0(\widetilde{t^*_L},\widetilde{t^*_R})\left(\begin{array}{cc}
     0~~~~ & 
     -(A^*_t c_\beta+\mu s_\beta)     \\[1mm]
     A_t c_\beta+\mu^* s_\beta~~~~ & 
     0
             \end{array}\right)
     \left(\begin{array}{c}
     \widetilde{t_L}        \\[1mm]
     \widetilde{t_R}       
             \end{array}\right)\,   \nonumber \\
&=&
\frac{m_t}{v\sin\beta}A^0(\widetilde{t^*_1},\widetilde{t^*_2})\left(\begin{array}{cc}
     2s_{\theta_{\widetilde{t}}}c_{\theta_{\widetilde{t}}}{\rm Im}[\hat{A_t}]~~~~ & 
     i(c^2_{\theta_{\widetilde{t}}}\hat{A_t}^*+s^2_{\theta_{\widetilde{t}}}\hat{A_t})     \\[1mm]
    -i(c^2_{\theta_{\widetilde{t}}}\hat{A_t}+s^2_{\theta_{\widetilde{t}}}\hat{A_t}^*)~~~~ & 
    -2s_{\theta_{\widetilde{t}}}c_{\theta_{\widetilde{t}}}{\rm Im}[\hat{A_t}]
             \end{array}\right)
     \left(\begin{array}{c}
     \widetilde{t_1}        \\[1mm]
     \widetilde{t_2}       
             \end{array}\right)\, \nonumber \\
&\equiv& A^0(\widetilde{t^*_1},\widetilde{t^*_2})\left(\begin{array}{cc}
     y^A_{\widetilde{t_1}\widetilde{t_1}}~~~~ & 
     {y^{A}_{\widetilde{t_1}\widetilde{t_2}}}^*     \\[1mm]
     y^A_{\widetilde{t_1}\widetilde{t_2}}~~~~ & 
     y^A_{\widetilde{t_2}\widetilde{t_2}}
             \end{array}\right)
     \left(\begin{array}{c}
     \widetilde{t_1}        \\[1mm]
     \widetilde{t_2}       
             \end{array}\right)
\end{eqnarray}
where $\hat{A_t}\equiv (A_t c_\beta+\mu^* s_\beta)e^{-i\delta_u}$,
and ${\rm Im}[\hat{A_t}]={\rm Im}[\mu^* e^{-i\delta_u}]/\sin\beta$.
Also,  $y^A_{\widetilde{t_1}\widetilde{t_1}} = -y^A_{\widetilde{t_2}\widetilde{t_2}}$.

The interaction between $Z$ boson and $\widetilde{t}_{L,R}$ is
\begin{eqnarray}
\mathcal{L}& \supset&
\frac{g}{\sqrt{1-x_W}} Z^{\mu} 
(\widetilde{t^*_L},\widetilde{t^*_R})  i\partials_\mu
\left(\begin{array}{cc}
     -\frac{1}{2}+Q_t x_W~~~~ & 
     0     \\[1mm]
     0~~~~ & 
     Q_t x_W
             \end{array}\right)
     \left(\begin{array}{c}
     \widetilde{t_L}        \\[1mm]
     \widetilde{t_R}       
             \end{array}\right)\,   \nonumber \\
&=&
\frac{g}{\sqrt{1-x_W}} Z^{\mu} \ 
(\widetilde{t^*_1},\widetilde{t^*_2}) i\partials_\mu
\left(\begin{array}{cc}
     -\frac{1}{2}c_{\theta_{\widetilde{t}}}+Q_t x_W   &
     \frac{1}{2} s_{\theta_{\widetilde{t}}}c_{\theta_{\widetilde{t}}}     \\[1mm]
     \frac{1}{2} s_{\theta_{\widetilde{t}}}c_{\theta_{\widetilde{t}}}~~~~ & 
     -\frac{1}{2}s^2_{\theta_{\widetilde{t}}}+Q_t x_W
             \end{array}\right)
     \left(\begin{array}{c}
     \widetilde{t_1}        \\[1mm]
     \widetilde{t_2}       
             \end{array}\right) \, \nonumber \\
&\equiv &
           Z^{\mu} (\widetilde{t^*_1},\widetilde{t^*_2})i\partials_\mu
\left(\begin{array}{cc}
     g^Z_{\widetilde{t_1}\widetilde{t_1}}~~~~ & 
     g^Z_{\widetilde{t_1}\widetilde{t_2}}     \\[1mm]
     g^Z_{\widetilde{t_1}\widetilde{t_2}}~~~~ & 
     g^Z_{\widetilde{t_2}\widetilde{t_2}}
             \end{array}\right)
     \left(\begin{array}{c}
     \widetilde{t_1}        \\[1mm]
     \widetilde{t_2}
                  \end{array}\right) \,,
\end{eqnarray}
where the two-way derivative $i\partials_\mu$ applies only to the stop
fields, and picks up  $(p-p')_\mu$ of the stop momenta $p,p'$ flowing into
the vertex in the Feynman diagram.

The process $\widetilde{t_1}\widetilde{t^*_1}\to hZ$ involve 
the $s$-channel diagram going by the $A^0$ exchange, as well as 
the $t$-channel and the conjugated $u$-channel  by  
the $\widetilde{t_2}$ exchange, as shown in Fig.~\ref{utsDiagram}.

In the non-relativistic approximation, the overall amplitude  is 
\begin{equation}
{\cal M}(\widetilde{t_1}\widetilde{t^*_1}\to hZ) =-\left[
\frac{4i {\rm Im}(g^{Z*}_{\widetilde{t_1}\widetilde{t_2}}y^h_{\widetilde{t_1}\widetilde{t_2}})}{m^2_h+m^2_Z-2(m^2_{\widetilde{t_1}}+m^2_{\widetilde{t_2}})}
+\frac{2y^A_{\widetilde{t_1}\widetilde{t_1}}g^Z_{Ah}}{4m^2_{\widetilde{t_1}}-m^2_A} \right]
 (P \cdot \varepsilon_Z)\,,
\end{equation}
where $g^Z_{Ah}=\frac{g}{2\sqrt{1-x_W}}\cos(\beta-\alpha)$.
The overall transition rate requires 
the polarization sum,
$$
\sum_{\varepsilon_Z}(P \cdot \varepsilon_Z)^2 
= P^\mu\left( -g_{\mu\nu}+\frac{p_{Z\mu}p_{Z\nu}}{m^2_Z} \right) P^\nu
=\frac{\lambda(s,m_h^2,m_Z^2)}{4m_Z^2}
%% =\frac{(s-m^2_h-m^2_Z)^2}{4m^2_Z}-m^2_h\,.
\ .$$
Here we use $2P\cdot p_Z={s+m^2_Z-m^2_h}$ and 
$s=m^2_{\widetilde\eta}\simeq 4m^2_{\widetilde{t_1}}$.
The kinematic function $\lambda(a,b,c)=a^2+b^2+c^2-2(ab+ac+bc)$.
Note that all amplitudes are suppressed by the non-alignment factor 
$\cos(\beta-\alpha)$,
which appears in both $g^Z_{Ah}$ and 
${\rm Im}[y^h_{\widetilde{t_1}\widetilde{t_2}}]
={\rm Im}[V_{LR}e^{-i\delta_u}]$.

The partial decay width in the non-relativistic approximation is
$$ \Gamma (\widetilde{t_1}\widetilde{t^*_1}\to hZ)=
\fr{1}{(2m_{\widetilde{t_1}})^2}
\sum_{\varepsilon_Z} |{\cal M}(\widetilde{t_1}\widetilde{t^*_1}\to hZ)|^2 
{|\psi(0)|^2} \fr3{8\pi}
\lambda^{1\over2}(1, m_h^2/s, m_Z^2/s)
\ ,$$
where the bound state wave function at the origin 
is estimated by the Coulomb type expression, 
$$ |\psi(0)|^2=\fr1{27\pi}(\alpha_s 2 m_{\widetilde t_1})^3 \ . $$
In comparison, we show the partial decay width of the gluon-gluon mode .
$$ \Gamma (\widetilde{t_1}\widetilde{t^*_1}\to gg)= 
\fr{4\pi\alpha_s^2}{3m_{\widetilde t_1}^2}|\psi(0)|^2 
\ .$$ 

\subsection{Contributions to the Electron EDM}

The most recent eEDM
gives a very stringent constraint\cite{Baron:2013eja}
\begin{equation}
|{d_e}|<8.7\times 10^{-29}~{e\cdot\rm [cm]} \ ,
\hbox{ at 90\% C.L.}    
\end{equation}
In MSSM, the relevant contribution to the eEDM 
based on the $CP$ violating parameters in the stop sector 
$\widetilde{t}_{1,2}$ 
arises via  the two-loop 
Barr-Zee diagrams~\cite{Chang:1998uc}.
\begin{equation}
\left( \frac{d_e}{e} \right)^{\widetilde{t}}_{\rm 2-loop}=2 Q_e Q^2_t
\frac{3 \alpha_{\rm em}}{64\pi^3}
\frac{ m_e}{m^2_A}
\left( 
    \frac{\sin2\theta_{\widetilde{t}}~m_t{\rm Im}[\mu^* e^{-i\delta_u}]}
     {v^2 \sin\beta\cos\beta}
\right)
  \left[ F\left(\frac{m^2_{\widetilde{t_1}}}{m^2_A}\right)
-F\left(\frac{m^2_{\widetilde{t_2}}}{m^2_A}\right) \right]\,, \label{13}
\end{equation}
where $\alpha_{\rm em}=e^2/(4\pi)$, 
$v \simeq 246$ GeV, 
and $F(z)$ is a two-loop function given by
\begin{equation}
F(z)=\int^1_0 dx \frac{x(1-x)}{z-x(1-x)}~{\rm ln}\left[ \frac{x(1-x)}{z} \right]\,.
\end{equation}

In fact the eEDM contribution vanishes in two different
limits, first when $A^0$ becomes heavy and decoupled, and second when
$m_{\widetilde t_1} \simeq m_{\widetilde t_2}$ so that their effects
cancel each other. Our numerical results show that even in 
general cases, an ample parameter space satisfies the eEDM constraint,
but still gives significant branching ratio mode of $hZ$.

Although the one-loop contributions to the eEDM 1-loop also exist in the
neutralino-selectron diagram, and the chargino-sneutrino diagram, they
involve totally different $CP$ violating parameters
and can be tuned to give tiny eEDM \cite{Ibrahim:1998je}. 
Therefore, we ignore their one-loop effect in eEDM.
In another approach \cite{Cheung:2014oaa,Bian:2014zka}, one can allow the sole
contribution of one type of diagrams to exceed the current experimental
limit, where one can expect that there might be other
types of diagrams that would cancel one another.

%%%%%%%%%%%%%%%%%%%%%%%%%%%%%%%%%%%%%%%%%%%%%%%%%%%%%%%%%%%%%%%%%%%%%%%%%%%%%

\section{Analysis}

The input parameters that are relevant for the stoponium decay into
$Zh$ are: 
$m_{\widetilde{t_1}}$, $m_{\widetilde{t_2}}$, Re$[\mu^* e^{-i\delta_u}]$, 
Im$[\mu^* e^{-i\delta_u}]$, $\theta_{\widetilde{t}}$, $\tan\beta$, and 
$m_A$. In the computation of the branching ratios of the stoponium,
it also involves the gluino mass $m_{\tilde{g}}$ and $\cos(\beta - \alpha)$.

Since we expect the pseudoscalar resonance can enhance the decay rate when 
$m_{\widetilde\eta}$ is around the heavy pseudoscalar $A^0$ mass,
we study the following cases, 
\begin{enumerate}
\item Near and below the pole, $m_{\widetilde\eta} < m_{A}$ 
by setting $2m_{\widetilde t_1}=1200$~GeV and $m_A=1.5$~TeV.

\item  Well below the pole, $m_{\widetilde\eta} \ll m_{A}$ 
by setting $2 m_{\widetilde t_1}=1200$~GeV and $m_A=2.5$~TeV.

\item Far from the pole for an extremely heavy $m_{A}$. 
We set $2 m_{\widetilde t_1}=1200$~GeV $\ll  m_A$.
In this case, we simply remove the $s$-pole contribution. Note that in this
limiting case, the two-loop contribution of the
pseudoscalar boson $A^0$ to the eEDM vanishes as well.
\end{enumerate}

\begin{figure}[t!]
\centering
\includegraphics[height=2.2in,angle=0]{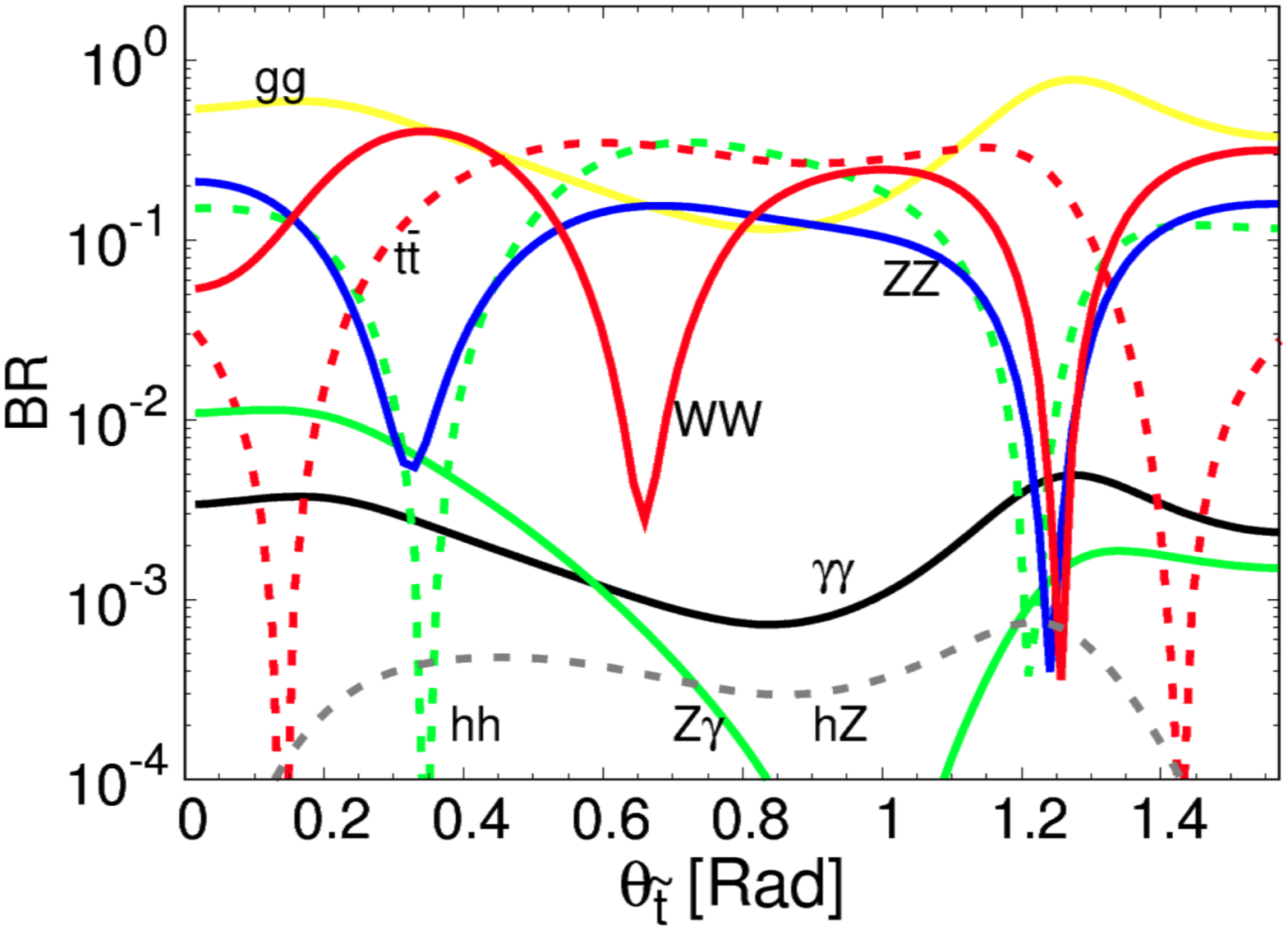}
\includegraphics[height=2.2in,angle=0]{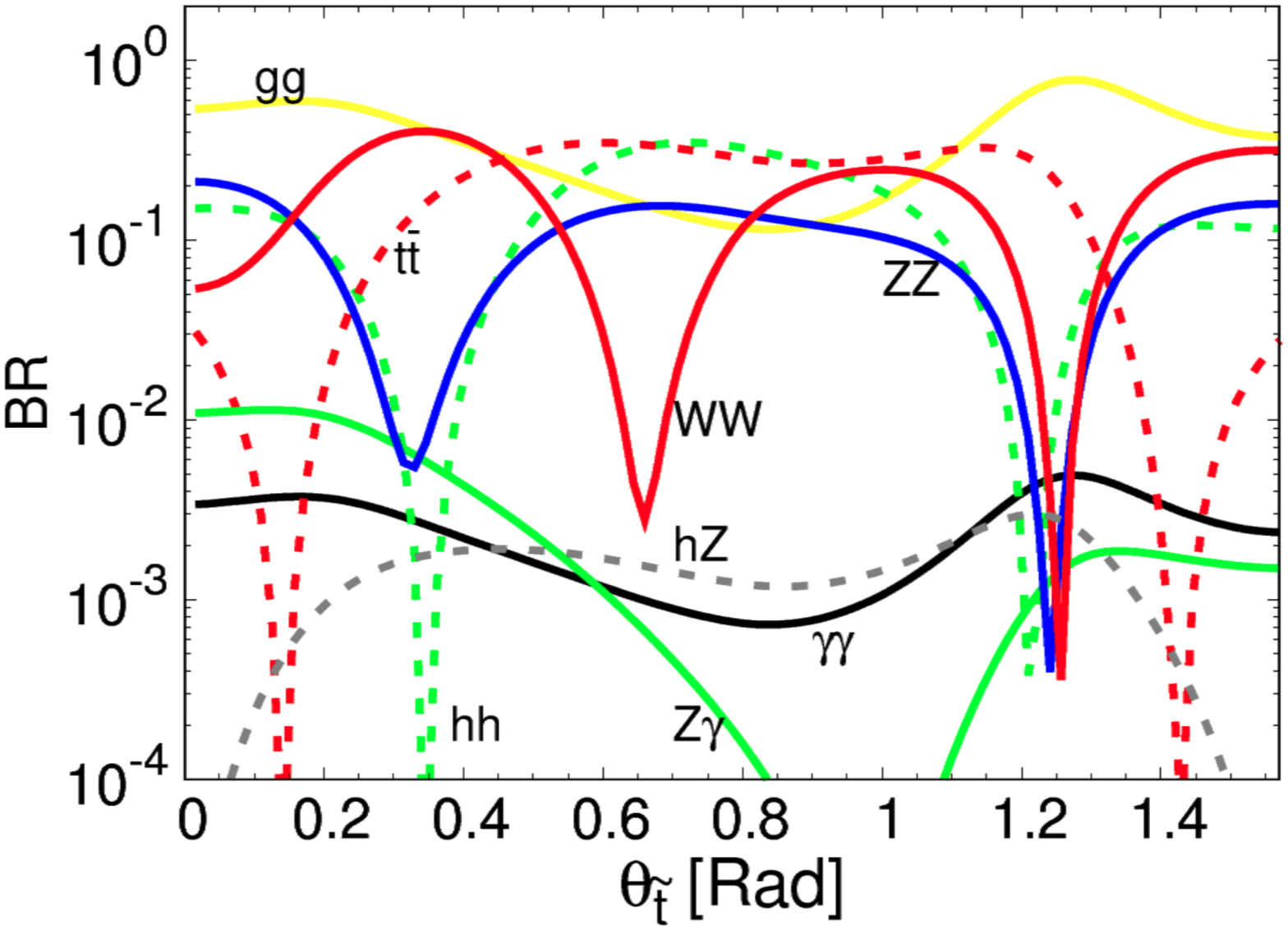}
\includegraphics[height=2.2in,angle=0]{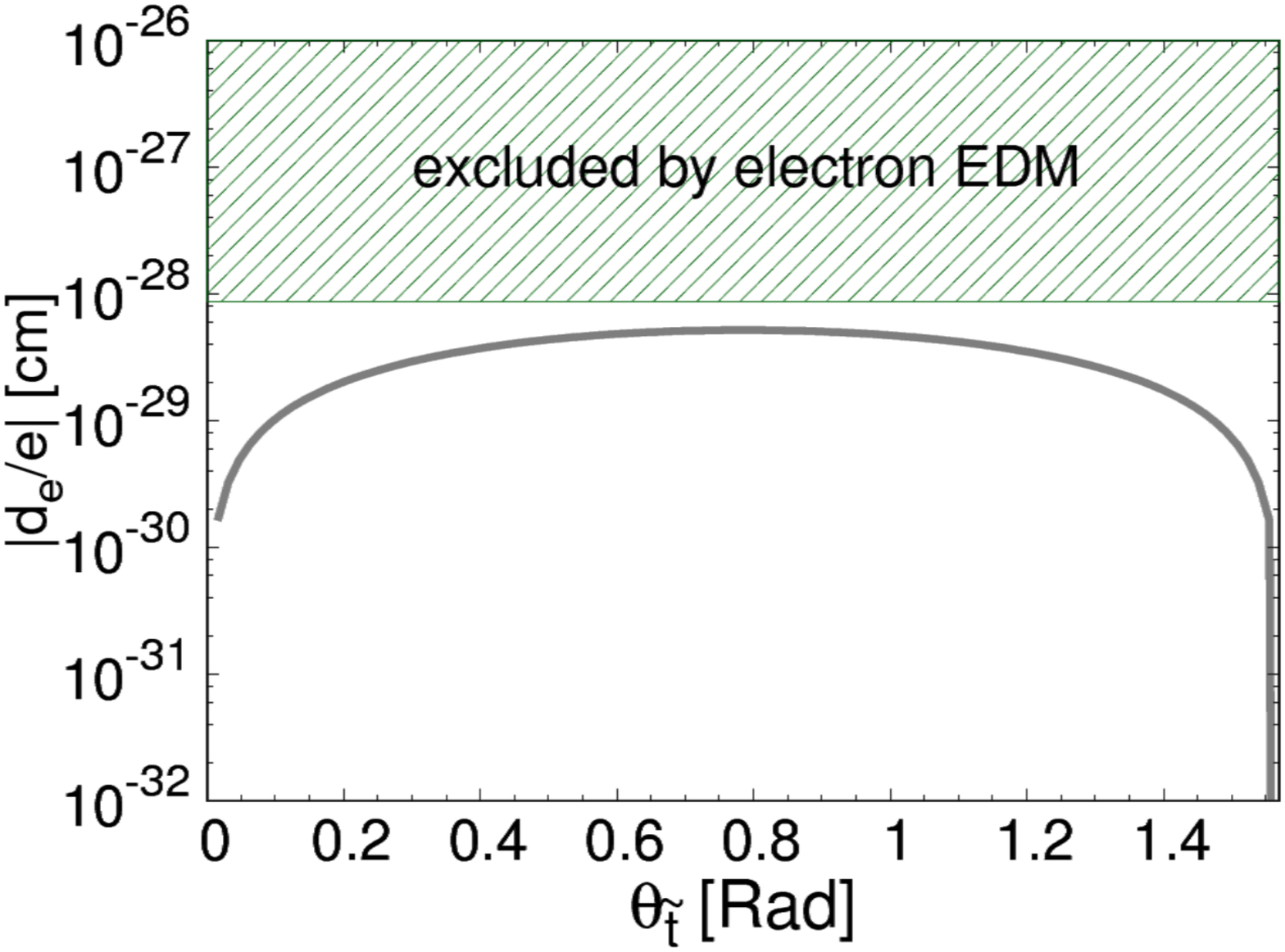}
\includegraphics[height=2.2in,angle=0]{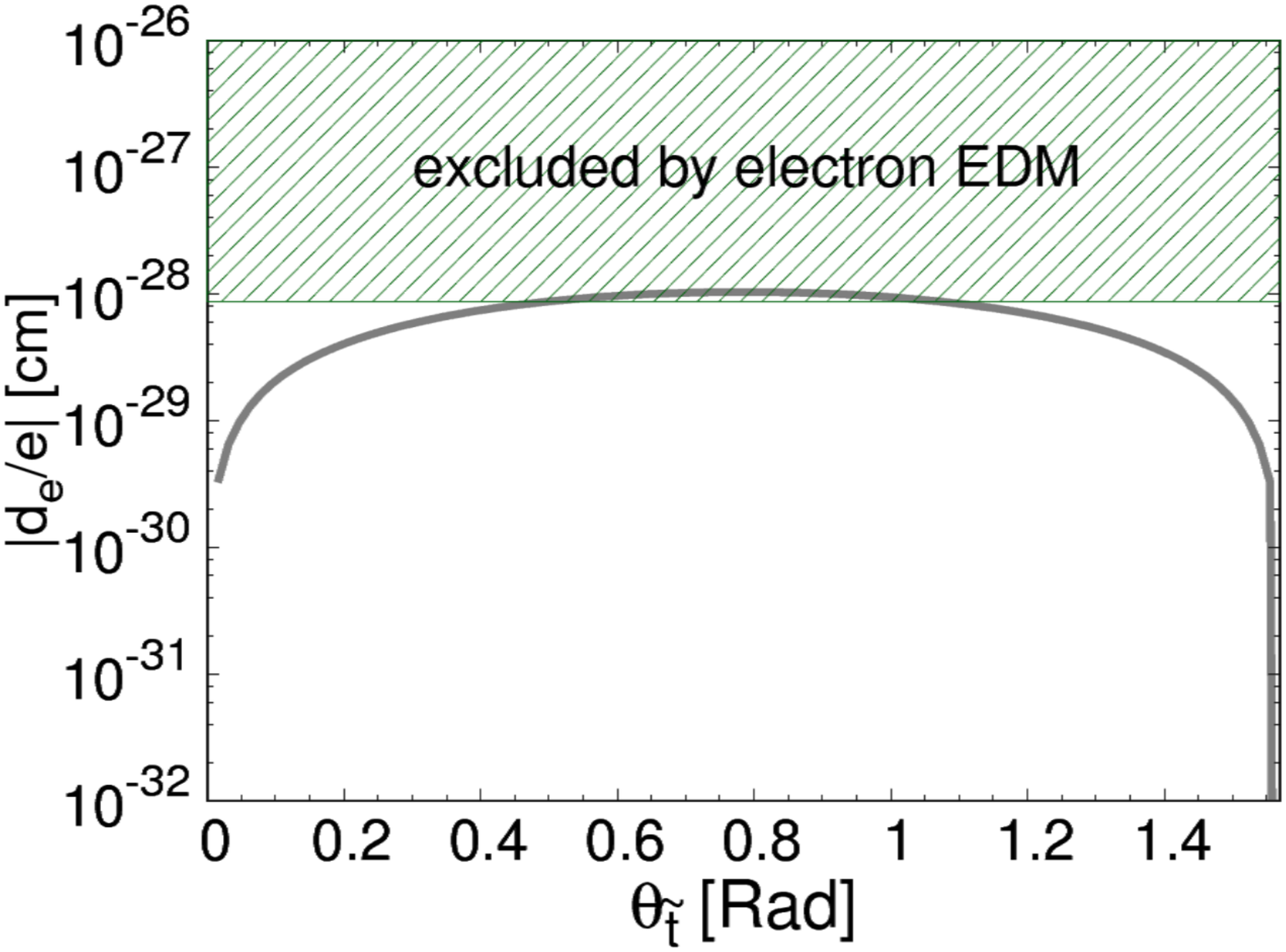}
\caption{\small \label{fig_lite_mu}
Upper panels show the branching ratios of the stoponium with
the corresponding predictions for the predicted eEDM from the two-loop 
Barr-Zee diagrams shown in the lower panels.
We set $\delta_u=0$, and $\mu$ purely imaginary, i.e. 
${\rm Re}[\mu^*e^{-i\delta_u}]=0$.
In the left (right) panel, we choose
${\rm Im}[\mu^*e^{-i\delta_u}]=100$~(200)~GeV.
For all panels we fix $m_{\widetilde{t}_1}=600$ GeV, $m_{\widetilde{t}_2}=1$ TeV,
$m_{\widetilde{g}}=2$ TeV, $\tan\beta=10$, $\cos(\beta-\alpha)=0.1$, 
$m_h=125$ GeV, $m_{H,A}=1.5$ TeV, and 
vary $\theta_{\widetilde{t}}\subseteq [0,\frac{\pi}{2}]$.
We include the binding energy effect in the stoponium mass, 
$m_{\widetilde\eta}=1195$ GeV.
}
\end{figure}

Note that we do not choose $m_A$ very close to $m_{\widetilde{\eta}}$
in case (1), because for such a low $m_A$ the contribution to the 
eEDM would be large. In Fig.~\ref{fig_lite_mu}, we show the 
branching ratios of the stoponium in upper panels with the corresponding
predictions for the eEDM in the lower panels, where we have chosen
the heavier stop mass $m_{\widetilde{t_2}}$ to be 1 TeV and $m_A = 1.5$ TeV.
For simplicity we have also chosen Re$[\mu^* e^{-i\delta_u}]=0$.
We note that the partial width into $Zh$ depends on Im$[\mu^* e^{-i\delta_u}]$,
as indicated in Eq.~(\ref{7}), and the eEDM is also proportional to
Im$[\mu^* e^{-i\delta_u}]$, as shown in Eq.~(\ref{13}).
Therefore, we cannot choose the parameter Im$[\mu^* e^{-i\delta_u}]$
arbitrarily large.  It is clear from the lower panels in 
Fig.~\ref{fig_lite_mu} that
Im$[\mu^* e^{-i\delta_u}]=200$ GeV is the largest allowed value without 
violating the constraint of eEDM under the set of other input parameters
shown in the figure caption. The branching ratio in $Zh$ is also small,
and of order $10^{-3}$ only.

\begin{figure}[t!]
\centering
\includegraphics[height=2.2in,angle=0]{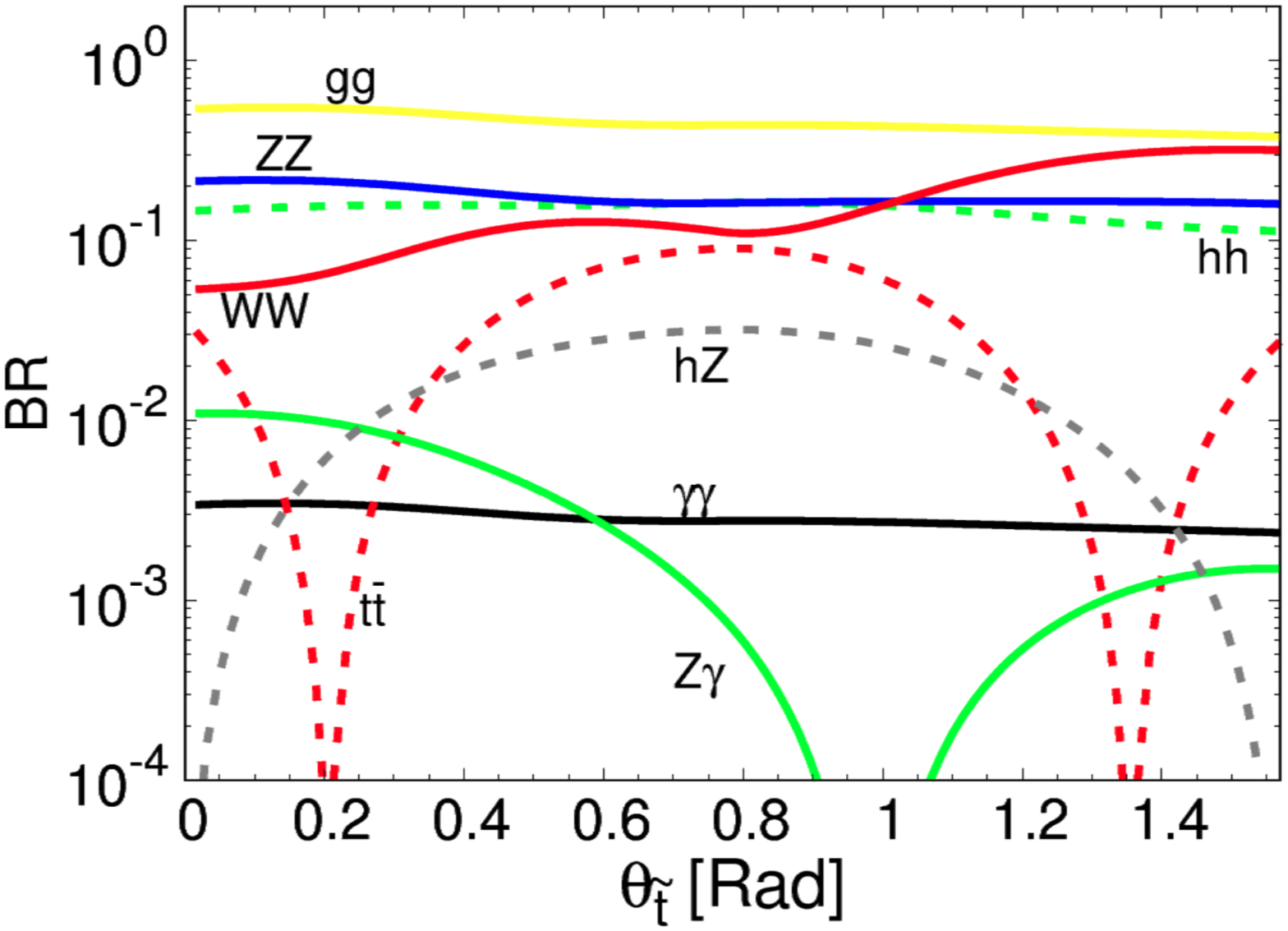}
\includegraphics[height=2.2in,angle=0]{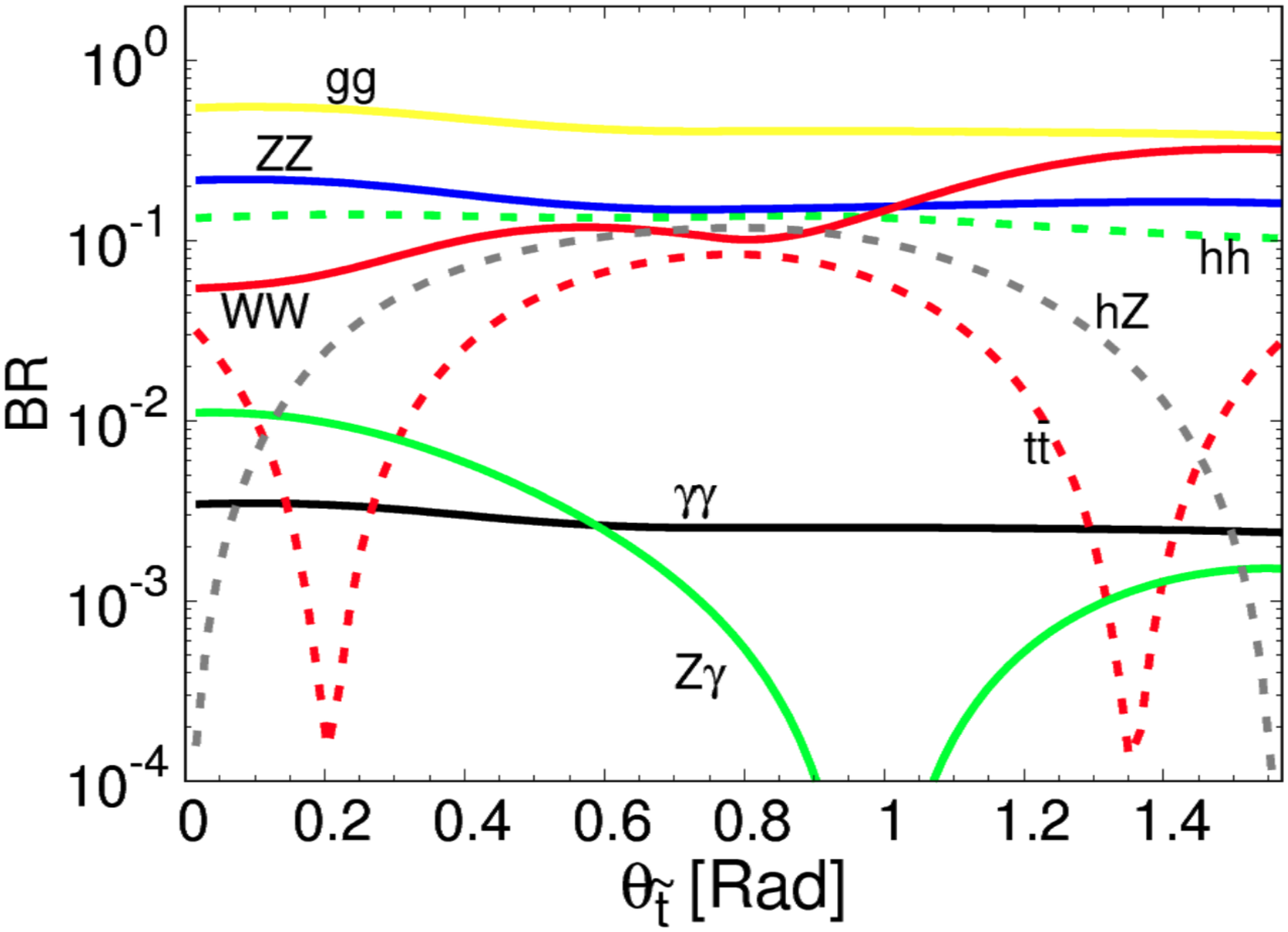}
\includegraphics[height=2.2in,angle=0]{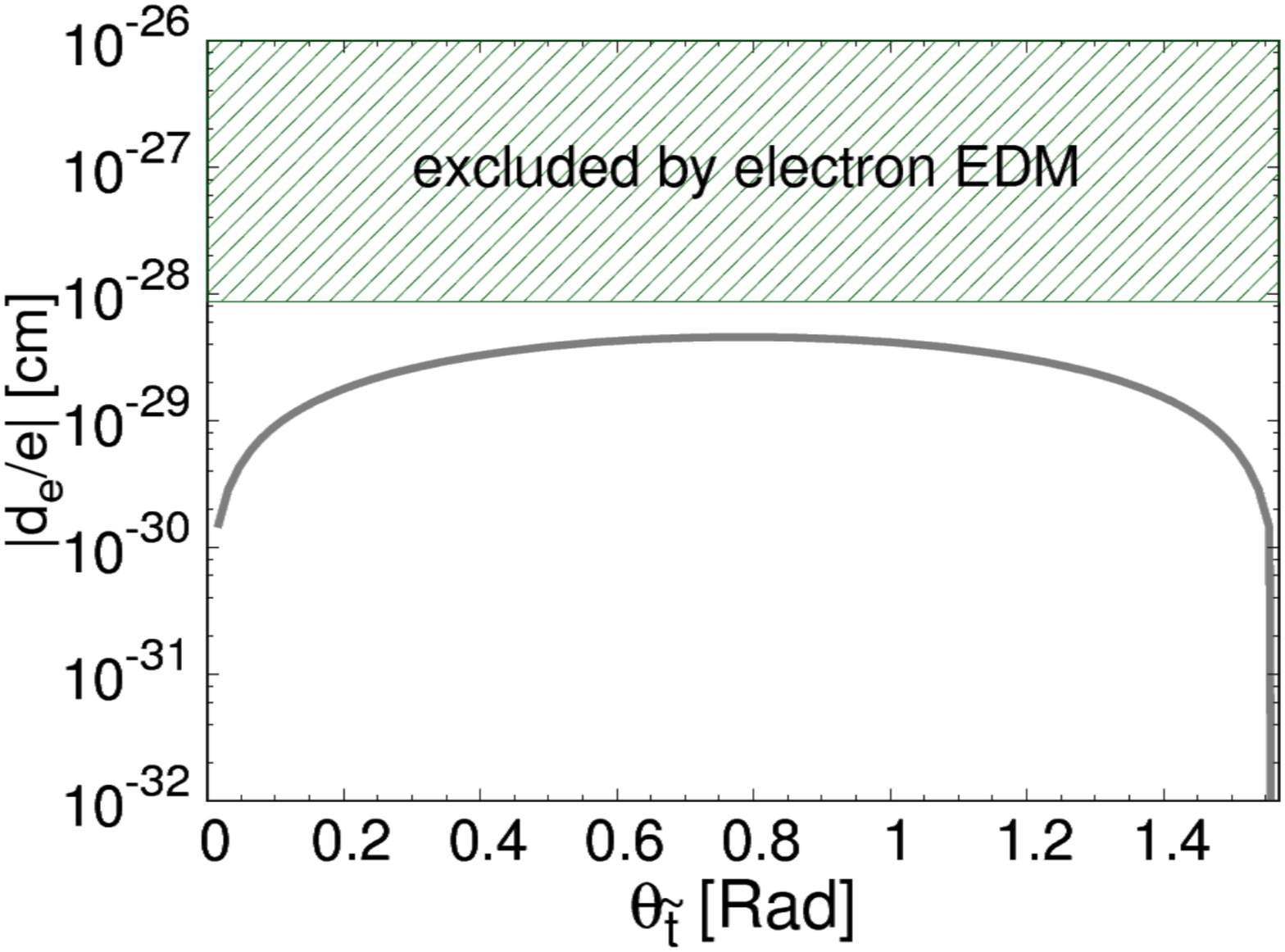}
\includegraphics[height=2.2in,angle=0]{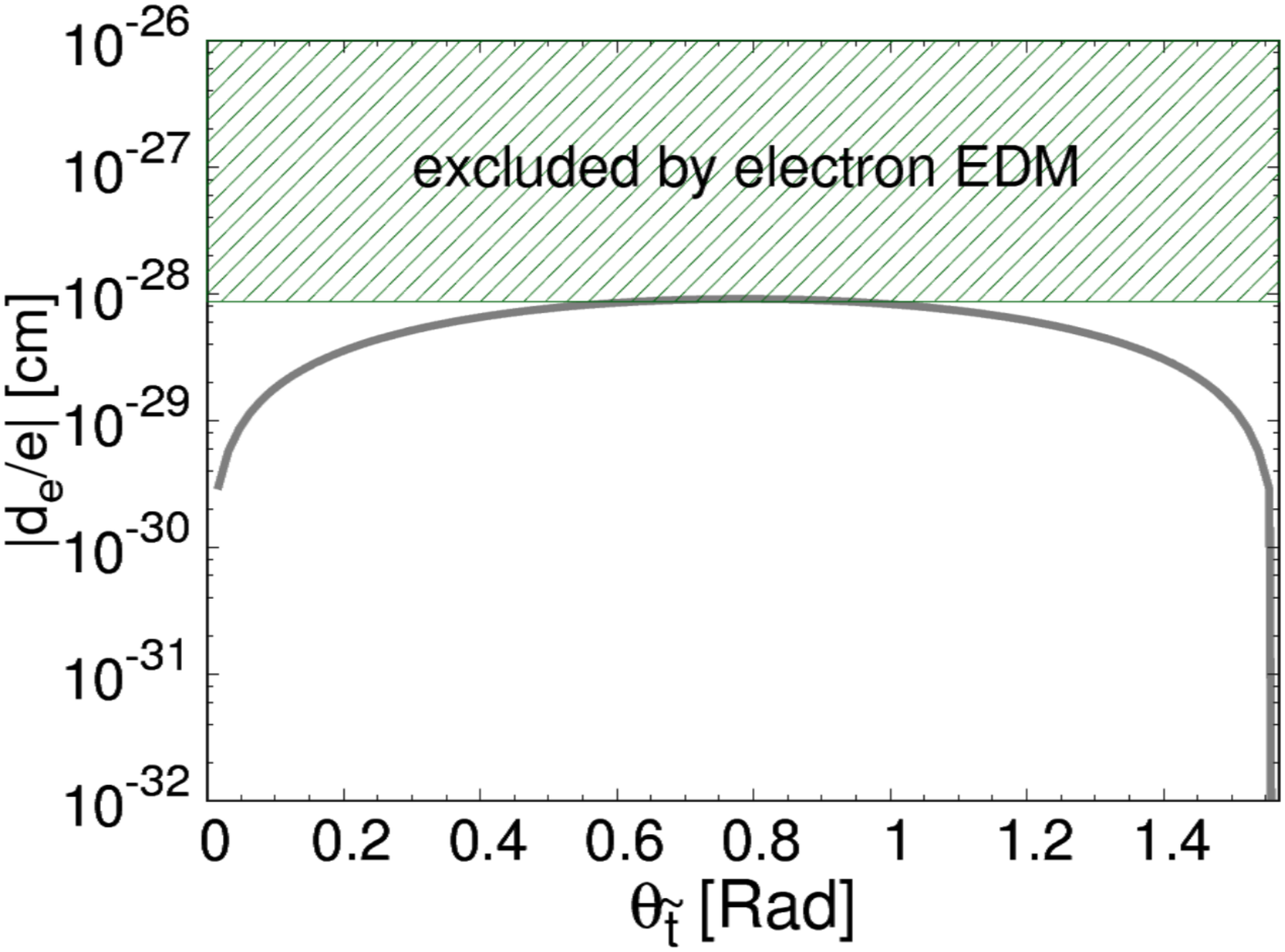}
\caption{\small   \label{fig_mid_mu} 
Upper panels show the branching ratios of the stoponium with
the corresponding predictions for the predicted eEDM from the two-loop 
Barr-Zee diagrams shown in the lower panels.
We set $\delta_u=0$, and $\mu$ purely imaginary, i.e. 
${\rm Re}[\mu^*e^{-i\delta_u}]=0$.
In contrast to Fig.~\ref{fig_lite_mu}, here we set $m_{H,A}=2.5$~TeV, 
$m_{\widetilde{t}_2}=650$~GeV, and 
${\rm Im}[\mu^*e^{-i\delta_u}]=1000$~(2000)~GeV  for the left (right) panels.
The other input parameters are the same as Fig.~\ref{fig_lite_mu}.
}
\end{figure}

An interesting observation can be found in Eq.~(\ref{13}) that when
the heavier stop mass is indeed close to the lightest stop mass, a 
significant cancellation between these two contributions is possible.
In Fig.~\ref{fig_mid_mu}, we show the 
branching ratios of the stoponium and the corresponding
predictions for the eEDM with $m_{\widetilde{t_2}} = 650$ GeV and
heavier $m_A = 2.5$ TeV (case 2). The parameter
Im$[\mu^* e^{-i\delta_u}]$ can be chosen as large as 2000 GeV without
violating the constraint of eEDM. With such a large 
Im$[\mu^* e^{-i\delta_u}]$ the branching ratio into $Zh$ can be as large 
as 10\%. With such a large branching ratio, the stoponium decay into
$Zh$ now becomes very interesting and detectable.

\begin{figure}[t!]
\centering
\includegraphics[height=2.2in,angle=0]{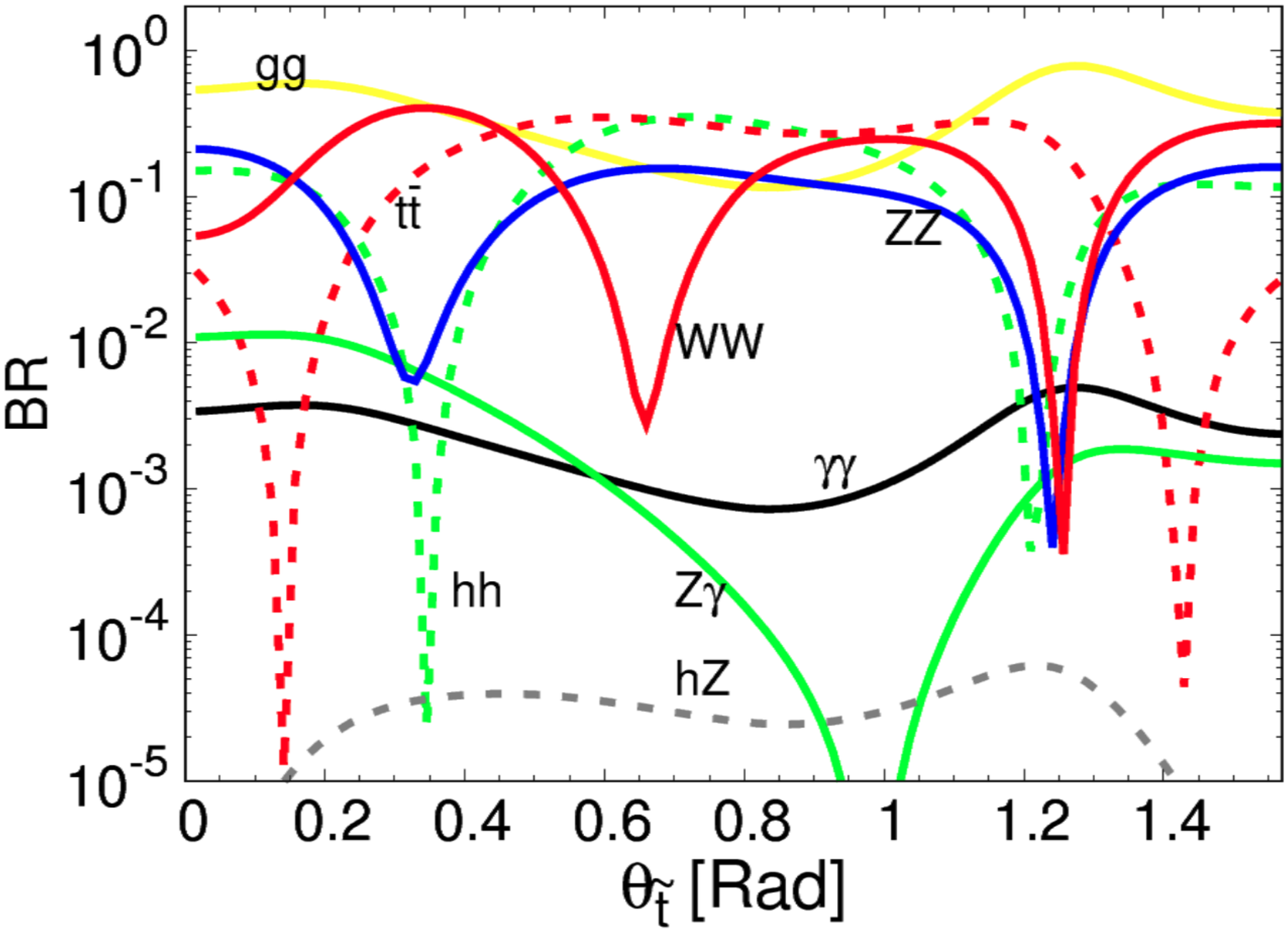}
\includegraphics[height=2.2in,angle=0]{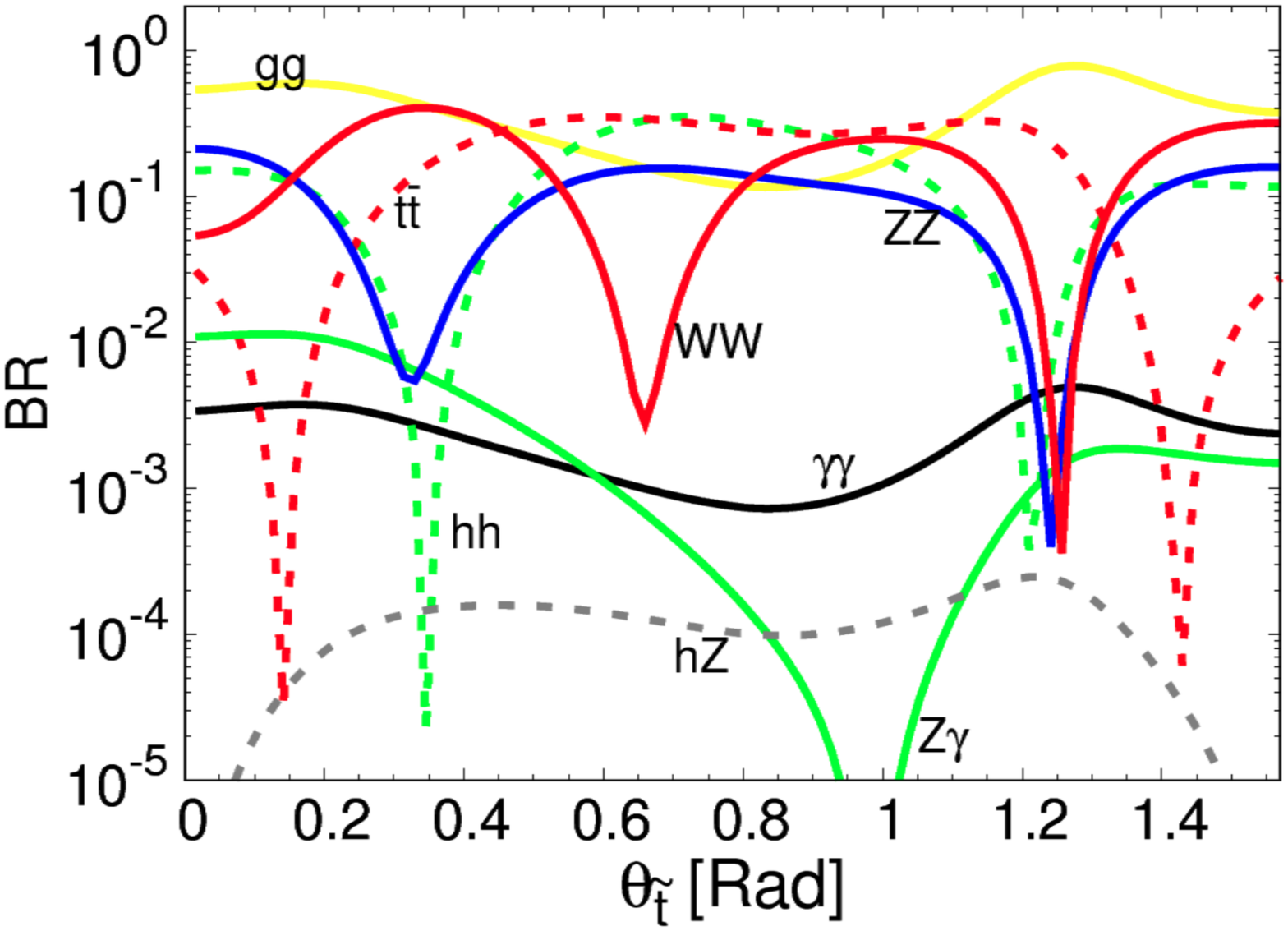}
\includegraphics[height=2.2in,angle=0]{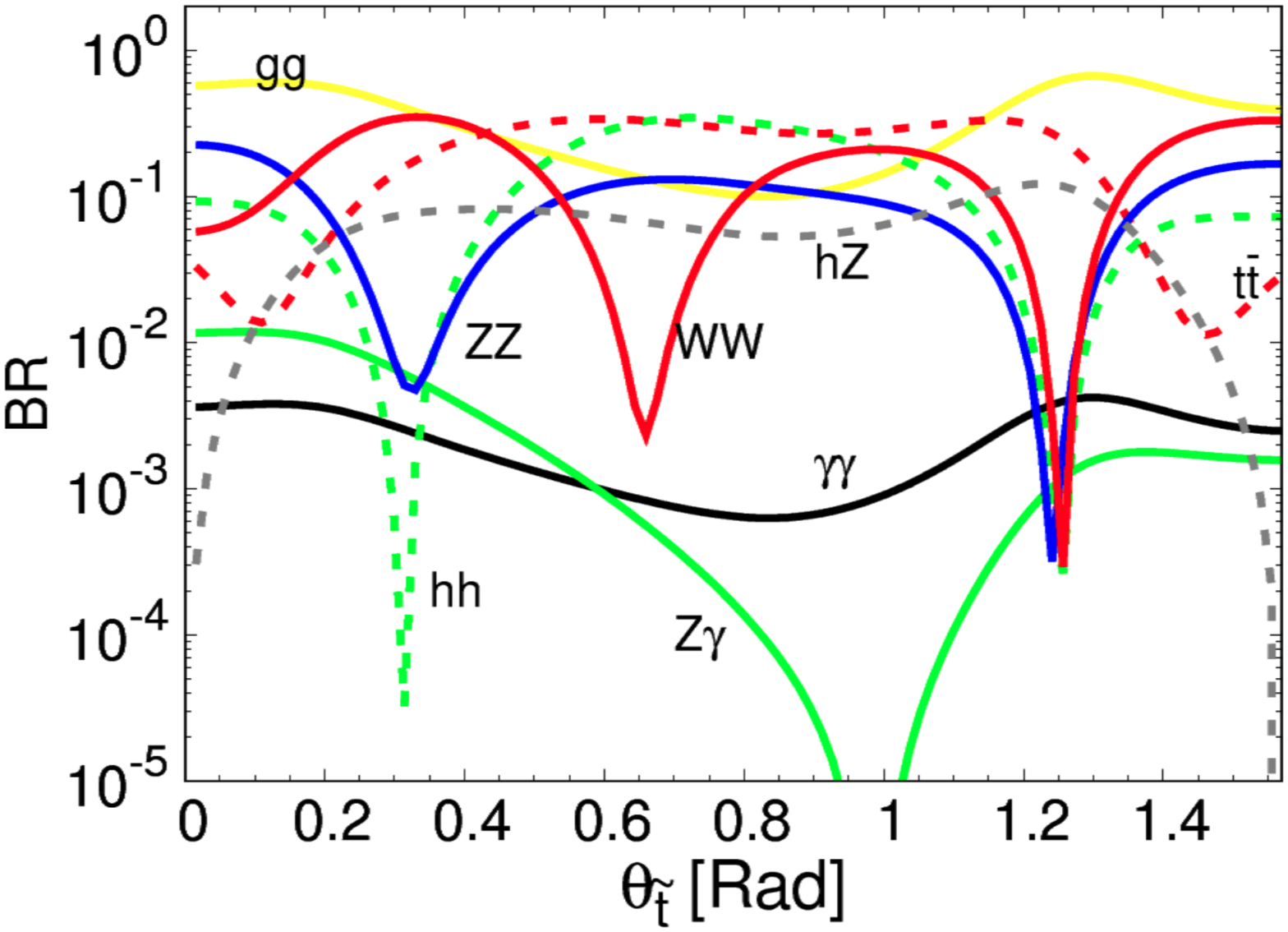}
\includegraphics[height=2.2in,angle=0]{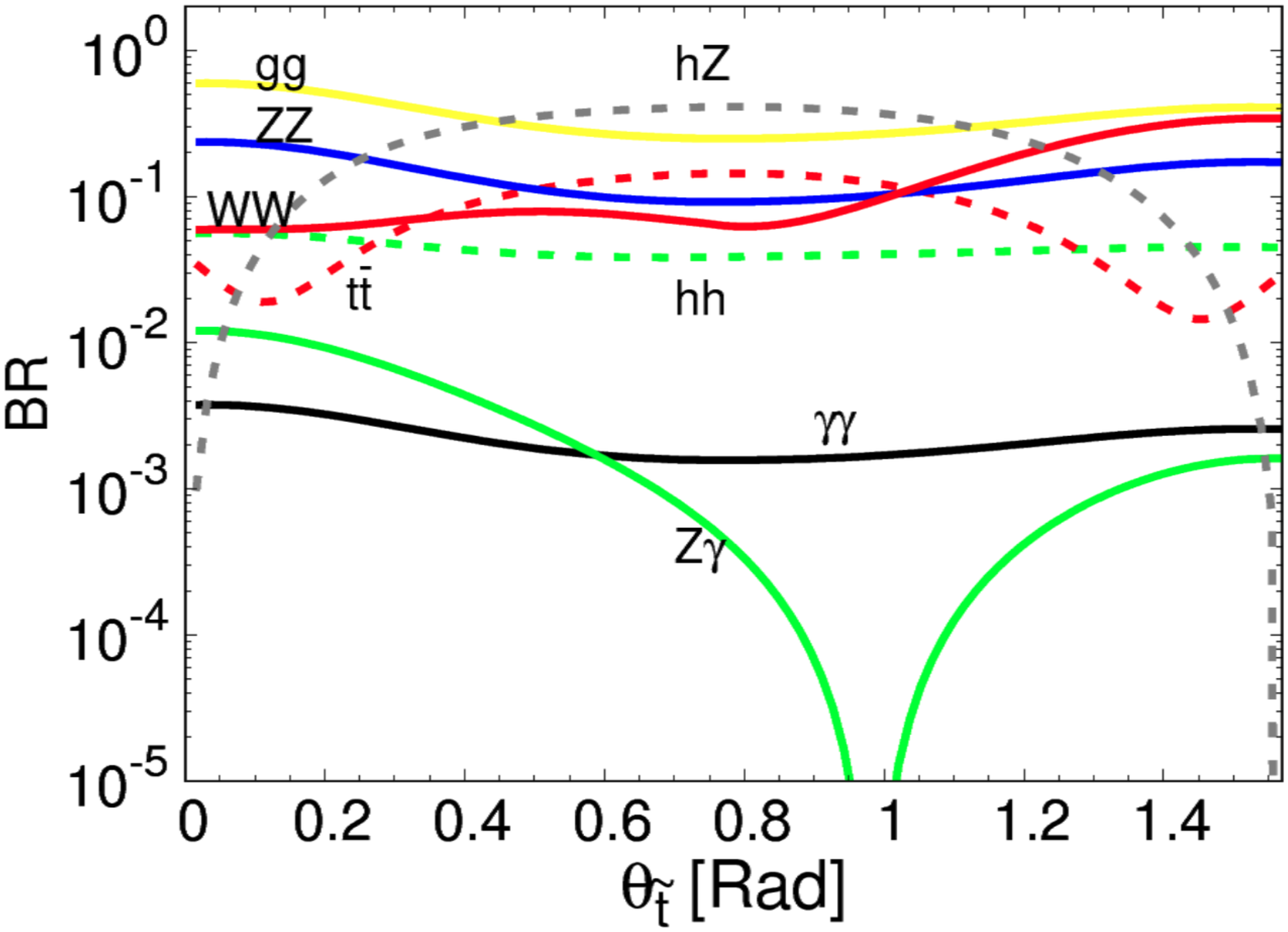}
\caption{\small   \label{fig_heav_mu}
In this extreme case of $m_A \to \infty$ the contribution to eEDM vanishes.
Here we show the branching ratios of the stoponium for 
$m_{\widetilde{t_1}}= 600$ GeV. The other relevant parameters are 
$m_{\widetilde{t}_2}=1$~TeV in the upper-left, upper-right, and lower-left panels,
while $m_{\widetilde{t}_2}=650$ GeV in the lower-right panel.
The ${\rm Im}[\mu^*e^{-i\delta_u}]=100,\,200$, $5000$, and 5000 GeV, 
respectively.
The other parameters are the same as in Fig.~\ref{fig_lite_mu}.
}
\end{figure}

In the extreme case of case (3), the mass of the pseudoscalar $A^0$
is set to be very heavy. Practically, we ignore the term involving the
$A^0$ exchange. We show in Fig.~\ref{fig_heav_mu} the branching ratios 
for the stoponium with $m_{\widetilde{t_1}} = 600$ GeV and 
$m_{\widetilde{t_2}} = 1000$ GeV, except for the lower-right panel 
where $m_{\widetilde{t_2}} = 650$ GeV.  Since there are no more 
$A^0$ contribution to the eEDM, we can set the parameter 
${\rm Im}[\mu^*e^{-i\delta_u}]$ large enough to achieve
a dominant branching ratio for the $Zh$ mode. We have chosen
${\rm Im}[\mu^*e^{-i\delta_u}]=100,\,200$, $5000$, and 5000 GeV, respectively.
Note that increasing ${\rm Im}[\mu^*e^{-i\delta_u}]$ will also increase the $hh$
mode, because the partial width $\Gamma({\widetilde\eta}\to hh)\propto
|y^h_{\widetilde{t_1}\widetilde{t_2}}|^2$, and 
$\Gamma({\rm  stoponium}\to hZ)\propto 
{\rm  Im}[y^h_{\widetilde{t_1}\widetilde{t_2}}]$.
In the most favorable case, the branching ratio into $Zh$ can be of order
$O(0.5)$, as indicated in the lower-right panel.

%%%%%%%%%%%%%%%%%%%%%%%%%%%%%%%%%%%%%%%%%%%%%%%%%%%%%%%%%%%%%%%%%%%%%%%%%%%%%

\section{Observability at the LHC}  

The leading order(LO) production process for $\widetilde{\eta}$ at LHC is 
through 
the gluon-gluon fusion, $gg \to \widetilde{t}_1\widetilde{t^*_1}$. 
The cross section section can be expressed 
in term of its gluonic decay width as~\cite{Martin:2008sv}
\begin{equation}
\sigma(pp \to \widetilde{\eta})=\frac{\pi^2}{8m^3_{\widetilde{\eta}}} 
\Gamma (\widetilde{t_1}\widetilde{t^*_1}\to gg)
\int^1_\tau dx \frac{\tau}{x}g(x,Q)g(\tau/x,Q)\,,
\end{equation}
where $g(x,Q)$ is the gluon parton distribution function, 
and $\tau\equiv m^2_{\widetilde{\eta}}/s$ with the center of mass energy of 
$pp$ collision $\sqrt{s}$. 
For the parton distribution function, we used CTEQ6~\cite{cteq} with 
the factorization scale $Q=m_{\widetilde{\eta}}$.
The K-factor, which is the 
ratio between the next leading order (NLO) and the LO  cross
sections, we take a reasonable value about 1.4. For more detailed NLO
calculation, we refer to Ref.~\cite{Younkin:2009zn}. At NLO, we obtain
the production cross section for 
$m_{\widetilde{\eta}}\simeq 1.2$~TeV at the LHC of  
$\sqrt{s}=13$~TeV.
\begin{equation}
\sigma(pp\to \widetilde{\eta})\simeq 1~[{\rm fb}]\,.
\end{equation}

%% Detectability of Zh resonance

The $Zh$ decay mode of the stoponium can be searched for via $h \to
b\bar b$ and $Z \to \ell^+ \ell^-$ or $Z \to jj$. At the LHC, such
searches have been performed
\cite{Aaboud:2017cxo,CMS:2018tuj,Aaboud:2017ahz, Aaboud:2018eoy}, in
which hadronic or leptonic modes of the $Z$ boson and $b\bar b$ mode
of the Higgs boson have been used. It is clear that the leptonic mode
of the $Z$ boson is clean but suffers from a small branching
ratio. The hadronic mode of $Z$ boson was believed to be suffered from
large QCD background. Nevertheless, with the advance of various
boosted-jet techniques the hadronic decays of the $Z$ boson and $h$
can be performed with reasonable success.  Since the stoponium is
rather heavy $\sim 1.2-1.5$ TeV here, the $Z$ boson and the Higgs boson
are very boosted with $p_T \sim 0.6 - 0.75$ TeV. The opening angle
between the decay products of the $Z$ or the Higgs boson is $\sim 2 M
/ p_T \sim 0.3 - 0.5$. This is in the right ballpark for excellent
detectability of boosted jets in contrast to the conventional QCD
background.

The recent search for $pp \to X \to Zh \to jj b\bar b $ performed by 
ATLAS\cite{Aaboud:2017ahz} at the LHC gave 
an upper limit on 
$\sigma(pp \to X \to Zh)\times B(h \to b\bar{b} + c\bar c)
< 20 - 30$ fb around the resonance mass $1.2-1.5$ TeV.
On the other hand, the search $pp \to X \to Zh \to \ell^+\ell^- b\bar b $ was
also performed \cite{Aaboud:2017cxo}. The upper limit on
$\sigma(pp \to X \to Zh)\times B(h \to b\bar{b} + c\bar c) < 10$ fb.
Note that these searches was designated for vector resonances.
In the same paper, they also gave
$\sigma(pp \to A \to Zh)\times B(h \to b\bar{b} ) < 10$ fb for 
$m_A \approx 1.2$ TeV.
Therefore, the production cross section of the stoponium times
the branching ratio into $Zh$ is well below the current limits
at the LHC.

With a project luminosity of 300 fb$^{-1}$ at the end of Run II, we 
can expect about 15 events for 
$\widetilde{\eta} \to Zh \to (jj,\ell\ell) + b\bar b$
for an optimistic branching ratio $B(\widetilde{\eta}\to Zh) \sim 10\%$. 
We emphasize again that in $CP$-conserving case the stoponium would not
decay into $Zh$, yet a small branching ratio into $Zh$ would signal
a violation of $CP$ symmetry.

\section{Conclusions}

We have demonstrated that the decay mode of the ground state of the
stoponium, $\widetilde\eta \to Zh$, can have a dominant or significant
branching ratio if we choose suitable $CP$ violating mixing in the
stop sector, which is still allowed by the eEDM measurement. Observation
of such a decay mode of the stoponium is clean signal of $CP$ violation.
The detailed
phenomenology will be investigated in a separate analysis.

Our framework for the decay mode $Zh$ from the scalar pair in the
ground state can be extended to other models that have fundamental
colored scalar bosons, such as the technipion\cite{Barger:1986bb} or 
the colored octet Higgs\cite{Gresham:2007ri}.

We offer a few comments before closing.
\begin{enumerate}
\item
Both the partial width of $\widetilde{\eta} \to Zh$ and eEDM increase
with increase in the parameter ${\rm Im}[\mu^*e^{-i\delta_u}]$. Therefore,
we cannot make it arbitrarily large. When $m_A =1.5$ TeV and 
$m_{\widetilde{t_1}} = 600$ GeV, ${\rm Im}[\mu^*e^{-i\delta_u}]$
can only be $100-200$ GeV.

\item The $A^0$ contribution would be suppressed with
increases in $m_A$. Further suppression can be achieved with
a smaller mass difference between $m_{\widetilde{t_1}}$ and $m_{\widetilde{t_2}}$.
For $m_{\widetilde{t_1}}=600$ GeV and $m_{\widetilde{t_2}}=650$ GeV, and 
$m_A = 2.5$ TeV, the parameter ${\rm Im}[\mu^*e^{-i\delta_u}]$ can be
as large as 2000 GeV. The branching ratio for $Zh$ can be enhanced to about
$0.1$.

\item In the extreme case of very heavy $m_A$, the $A^0$ contribution to
eEDM vanished. Thus, we can choose a very large ${\rm Im}[\mu^*e^{-i\delta_u}]$
such that the branching ratio into $Zh$ can be of order $O(0.5)$.

\item There are other contributions to the eEDM from 1-loop diagrams in
supersymmetric models, such as chargino-selectron loop and 
neutralino-sneutrino loop, and other 2-loop diagrams such as Barr-Zee 
diagrams with chargino, neutralino, stau, etc. Here in this work we 
only focus on a particular contribution from $A^0$. In principle, we 
can allow some level of cancellation from other contributions, such 
that the sole contribution from $A^0$ may be over the current constraint. 
In such a case, the parameter ${\rm Im}[\mu^*e^{-i\delta_u}]$
could be chosen a larger value and the branching ration into $Zh$ could 
increase.
\end{enumerate}

\section{\bf Acknowledgments}
W.-Y. K. and P.-Y. T. thank the National Center of Theoretical Sciences and
Academia Sinica, Taiwan, R.O.C. for hospitality.  This research was
supported in parts by the Ministry of Science and Technology (MOST) of
Taiwan under Grant Nos. MOST-105-2112-M-007-028-MY3,  and 
by the World Premier International Research Center Initiative (WPI), 
MEXT, Japan.

%% \newpage

%%%%%%%%%%%%%%%%%%%%%%%%%%%%%%%%%%%%%%%%%%%%%%%%%%%%%%%%%%

\end{document}